\documentclass[epj,twocolumn]{webofc}
\usepackage[varg]{txfonts}   
\usepackage{hyperref}
\usepackage{lineno}
\begin{document}

\woctitle{ISVHECRI 2018}

\title{{Recent results from the Pierre Auger Observatory}

\author{\firstname{Sergio} \lastname{Petrera}\inst{1,2}\fnsep\thanks{\email{sergio.petrera@aquila.infn.it}} 
  for the Pierre Auger Collaboration\inst{3}\fnsep\thanks{\email{auger_spokespersons@fnal.gov}}}}

\institute{Gran Sasso Science Institute (GSSI), L'Aquila, Italy
\and
           INFN Laboratori Nazionali del Gran Sasso, Assergi (L'Aquila), Italy  
\and
Observatorio Pierre Auger, Av. San Martín Norte 304, 5613 Malarg\"ue, Argentina
\\
Full author list: {\href{http://www.auger.org/archive/authors\_2018\_08.html}
{{http://www.auger.org/archive/authors\_2018\_05.html}}}}

\abstract{In this paper some recent results from
  the Pierre Auger Collaboration are presented. These are the measurement
of the energy spectrum of cosmic rays over a wide range of energies ($10^{17.5}$ to above $10^{20}$ eV),
studies of the cosmic-ray mass composition with the fluorescence and surface
detector of the Observatory,
the observation of a large-scale anisotropy in the arrival direction of cosmic rays above
$8 \times 10^{18}$ eV and indications of anisotropy at intermediate angular scales above $4 \times 10^{19}$ eV. The astrophysical implications of the spectrum and
composition results are also discussed. Finally the progress of the upgrade of the Observatory, AugerPrime is presented.
}
\maketitle
\section{Introduction}
\label{intro}

The Pierre Auger Observatory \cite{ThePierreAuger:2015rma} is the largest facility to detect cosmic rays built so far.
It is located in the province of Mendoza,
Argentina and has been in operation since 2004. Cosmic rays are studied by combining a 
Surface Detector (SD) and a Fluorescence Detector (FD) to measure extensive air showers.
The SD consists of 1600 water-Cherenkov detectors on a 1500 m triangular grid (SD-1500) over an area of about 3000 km$^2$, and of additional 61 detectors covering 23.5 km$^2$ on a 750 m grid (SD-750 or `infill' array).
The 24 fluorescence telescopes grouped in 4 FD buildings are located on the boundary of the observatory to overlook the whole atmospheric volume above the surface array. Three additional telescopes pointing at higher elevations (HEAT) are located near one of the FD sites (Coihueco) to detect lower energy showers.
An array of radio antennas, Auger Engineering Radio Array (AERA) \cite{Abreu:2012pi,ThePierreAuger:2015rma}, complements the data with the detection of the shower radiation in the hundred MHz region.

The design of the observatory has been conceived to exploit the ``hybrid'' concept, the simultaneneous detection of air showers by the surface array and fluorescence telescopes. The apparatus collects shower events of different classes depending on the on-time (generally called duty cycle) of the different detector components: the surface array is able to collect showers at any time, whereas the fluorescence detectors can operate only during clear moonless nights ($\approx$ 15\% duty cycle).  Taking into account geometry and quality cuts applied at the event reconstruction level, the common data-set is only few percent. Therefore only a small part of the SD showers are actually reconstructed by the FD. Nonetheless this sub-sample (the hybrid data-set) is very valuable, including events having both the footprint of the shower at ground and the longitudinal profile measured.  The hybrid approach has been a major breakthrough in the detection of UHECRs since the method allows one to have the same energy scale in the surface detectors and the fluorescence telescopes and to derive the energy spectra entirely data-driven and free of model-dependent assumptions about hadronic interactions in air showers

In this paper I summarize some recent results from the Pierre Auger Observatory. Other interesting outcomes on several aspects of cosmic ray and particle physics that are not included here, but can be found e.g. in \cite{Aab:2017njo} \cite{ObservatoryMichaelUngerforthePierreAuger:2017fhr}.

\begin{figure*}[t!]
    \centering
    \begin{tabular}{ll}
      \includegraphics[scale=0.20]{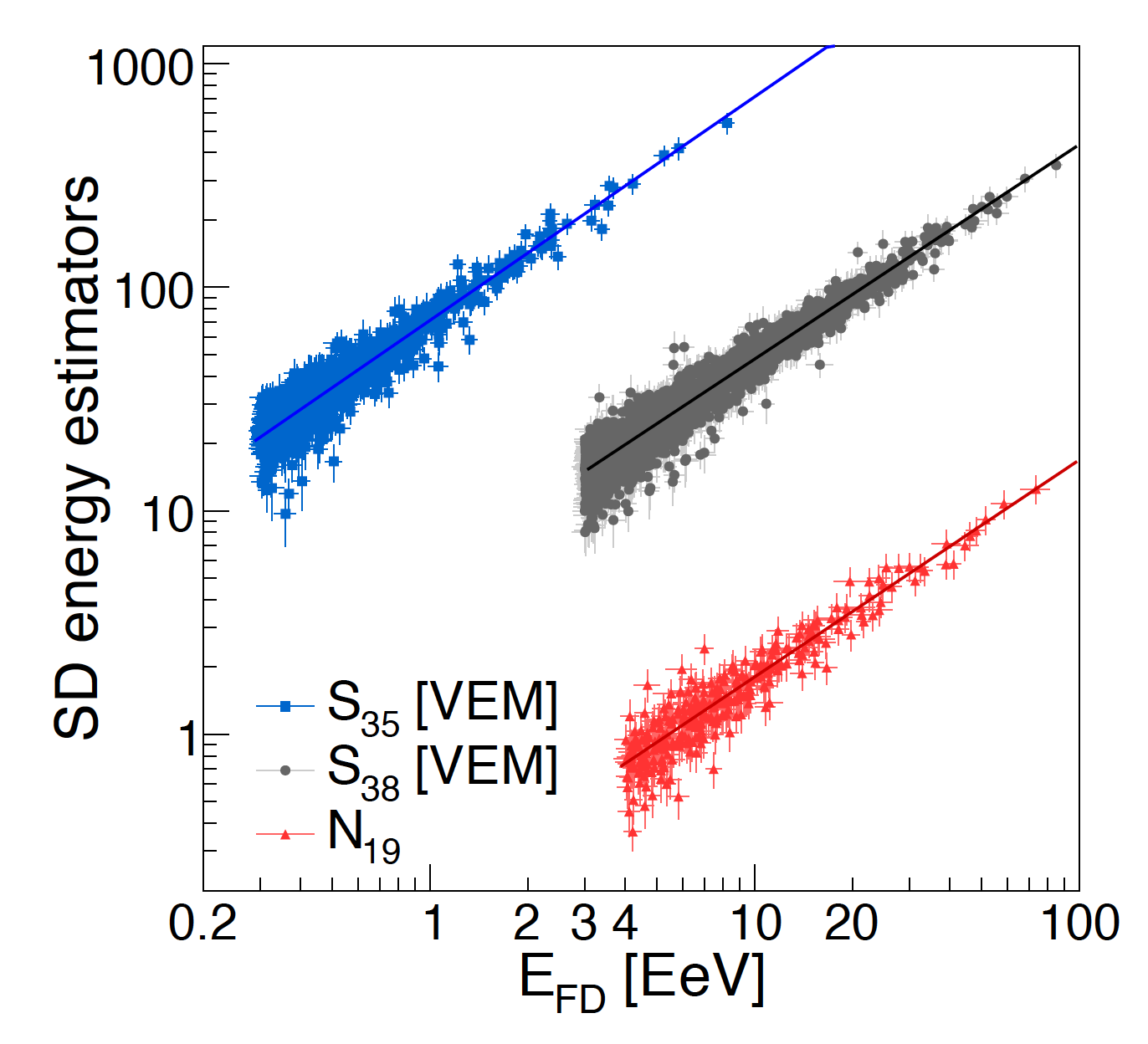}
      &  \includegraphics[scale=0.24]{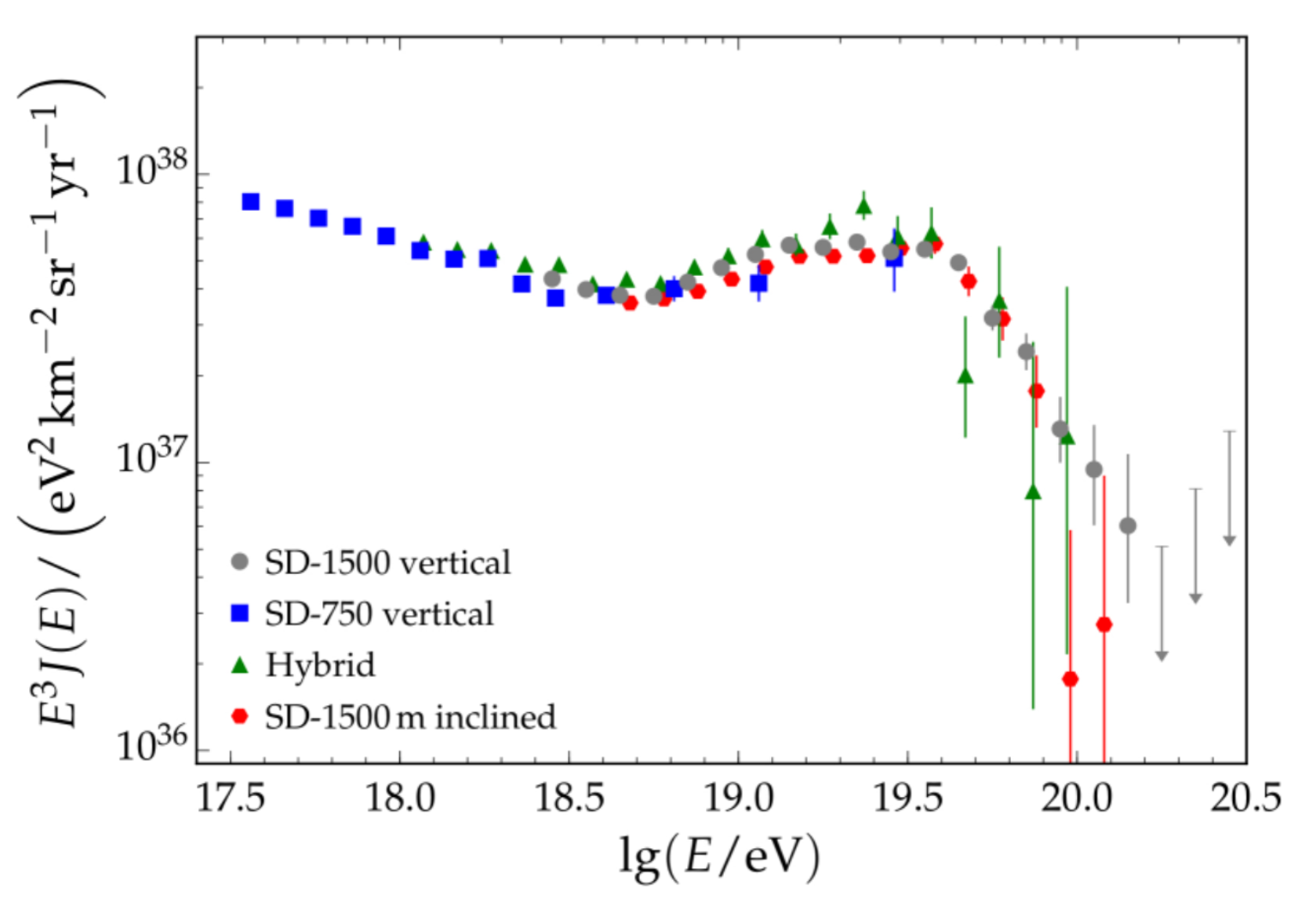}
    \end{tabular}
\caption{{\it Left:} Energy calibration of the surface detector. The shower size measured
  for `vertical' events with the SD-1500 ($S_{38}$) and SD-750 ($S_{35}$) array and for inclined
  showers ($N_{19}$) is shown as a function of the
  energy measured with the fluorescence telescopes ($E_\mathrm{FD}$). {\it Right:}
  The energy spectra obtained with the four spectrum components.
The systematic uncertainty on the energy scale, common
to all of them, is 14\%}
\label{fig:Spectra}
\end{figure*}

\section{Energy spectrum}
\label{sec:spectrum}
One of the main goals of the Pierre Auger Observatory is to measure the cosmic ray spectrum
at its highest energy end with unprecedented precision.
This is done exploiting all data collected
by the observatory, divided into 4 different samples. The SD-1500 and SD-750 samples are
made of `vertical' events (zenith angle $\theta < 60^\circ$ for the standard and $\theta < 55^\circ$ for the infill array)
observed by the standard SD array and the infill array respectively. The `inclined' sample is measured
with the standard SD array but covers different zenith angles ($60^\circ < \theta < 80^\circ$ ) where a different reconstruction is needed. Finally the hybrid sample is made of events observed by both SD and FD.

The analysis method to derive the spectra is entirely data-driven. For the surface detector data, we transform
the measured shower sizes to a size estimator that is independent of zenith angle by using the
method of constant intensities \cite{Hersil:1961zz} for the `vertical' data and templates of the footprint of the
particle densities at ground for the `inclined' data set \cite{Aab:2014gua}. These attenuation-corrected shower sizes
are used as energy estimates after calibrating them with the calorimetric energy available
for hybrid events as shown in the left panel of Fig. \ref{fig:Spectra}. Following
this method all the spectrum components have the same energy scale.
The overall systematic uncertainty of the energy scale remains at 14\% \cite{Aab:2017njo,Fenu:2017hlc,verziICRC13}.

\begin{figure}[t!]
\begin{center}
  \includegraphics[scale=0.23]{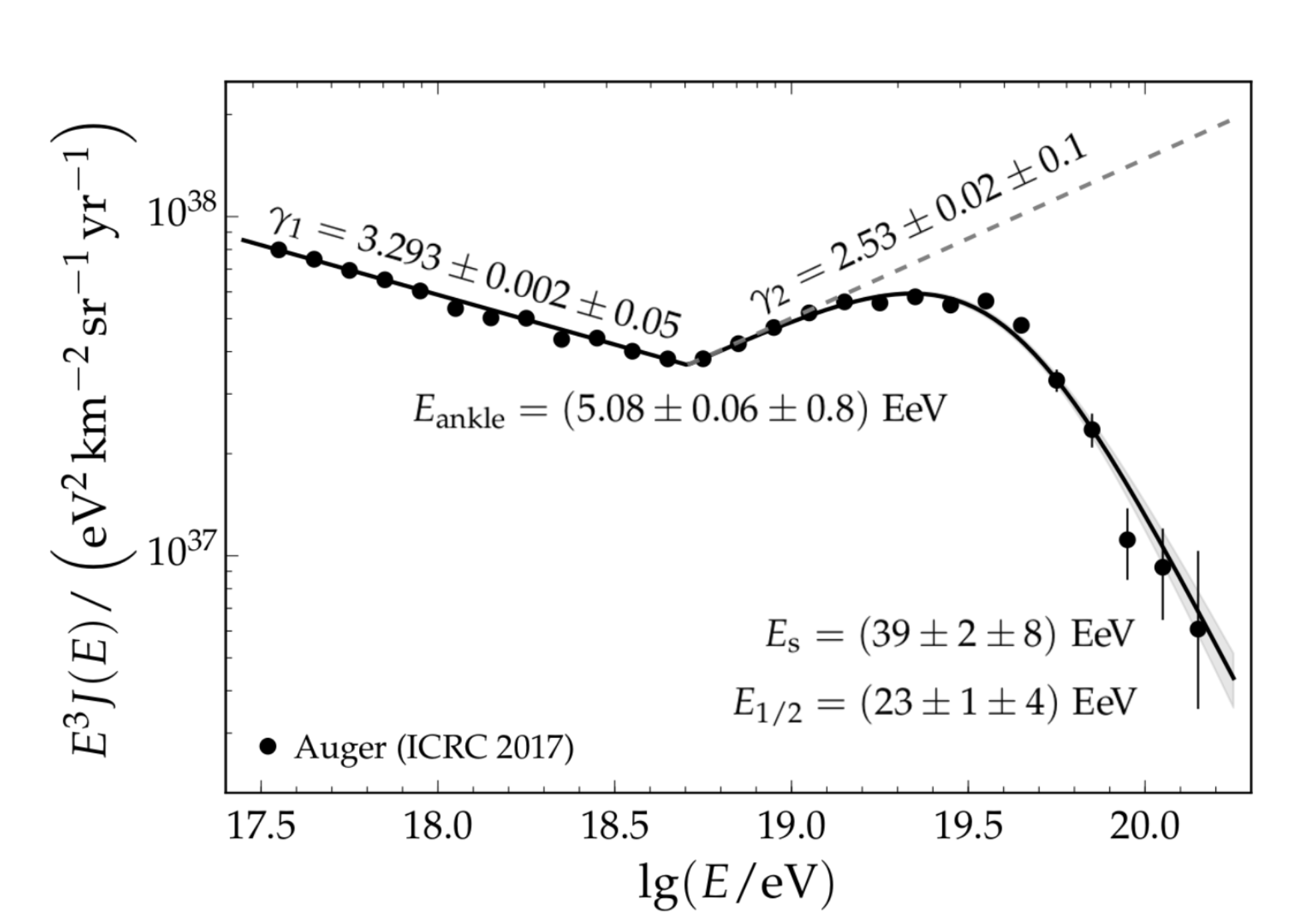}
 \end{center}
\caption{Combined energy spectrum. The line
shows a fit to the spectrum with a broken power law and a suppression at ultrahigh energies. The gray dashed
line indicates the same broken power law without suppression. The fitted spectral indices and energies of
the break and suppression are superimposed together with their statistical and systematic uncertainties.}
\label{fig:Features}
\end{figure}

All the spectra agree within the systematic uncertainties as shown in the
right panel of Fig. \ref{fig:Spectra}, and are combined through a maximum
likelihood fit in order to obtain the final spectrum.
The combined energy spectrum \cite{Aab:2017njo,Fenu:2017hlc} is shown in Fig. \ref{fig:Features} as presented at ICRC 2017. At the ``ankle'', observed at
$E_\mathrm{ankle} = 5.08 \pm 0.06 \mathrm{(stat.)} \pm 0.8 \mathrm{(syst.) ~EeV}$, the spectral index hardens by $\Delta \gamma \sim -0.76$.
A power law extension of the flux above the ankle is clearly excluded by data and we find a suppression energy\footnote{
We fitted the flux with a power law allowing for a break in the spectral
index at $E_\mathrm{ankle}$ and a suppression of the flux at ultrahigh energies
$\propto [1 + (E/E_s)^{\Delta \gamma_s}]^{-1}$
 } of $E_{s} = 39 \pm 2 \mathrm{(stat.)} \pm 8 \mathrm{(syst.) ~EeV}$ with a spectral index softening of $\Delta \gamma_s \sim 2.5$.
The energy at which the integral flux drops by a factor two below
what would be expected without suppression is found to be $E_{1/2} = 23 \pm 1 \mathrm{(stat)} \pm 4 \mathrm{(syst)~ EeV}$. This value is considerably lower than $E_{1/2} = 53$ EeV as predicted for the classical GZK scenario \cite{Berezinsky:2002nc} where the suppression at ultrahigh energies is caused by the propagation of extra-galactic protons.
However the suppression of the spectrum can also be described by
assuming a mixed composition at the sources or  by the limiting
acceleration energy at the sources rather than by the GZK-effect. Hence the energy spectrum alone remains ambiguous concerning astrophysical scenarios,
which are better studied complementing the spectrum with other CR observables like mass composition and anistropy.

Comparing this energy spectrum with the one by Telescope Array one finds \cite{AbuZayyad:2018aua} that the ankle energies are consistent within the systematic uncertainties in the energy scale, but the discrepancy between the cut-off energies is not explained by systematics.
 An interesting question is whether the cutoff energy difference is due to a systematic bias or to astrophysics. A possible contribution to this difference in terms of declination dependence of the flux, as suggested by TA \cite{AbuZayyad:2018aua}, has been investigated by Auger measuring the flux with the SD in different declination bands. No significant variation has been found that could account for the discrepancy between spectra measured from different hemispheres.  The differences found between the flux measured in two separate declination bands, `southern' (`northern'), corresponding to $\delta_\mathrm{d} < 29.47^\circ$ ($\delta_\mathrm{d} > 29.47^\circ$), are instead compatible with the variations expected from a dipolar modulation of the flux \cite{Valino:2015zdi}.

\begin{figure*}[t]
\begin{center}
\includegraphics[scale=0.4]{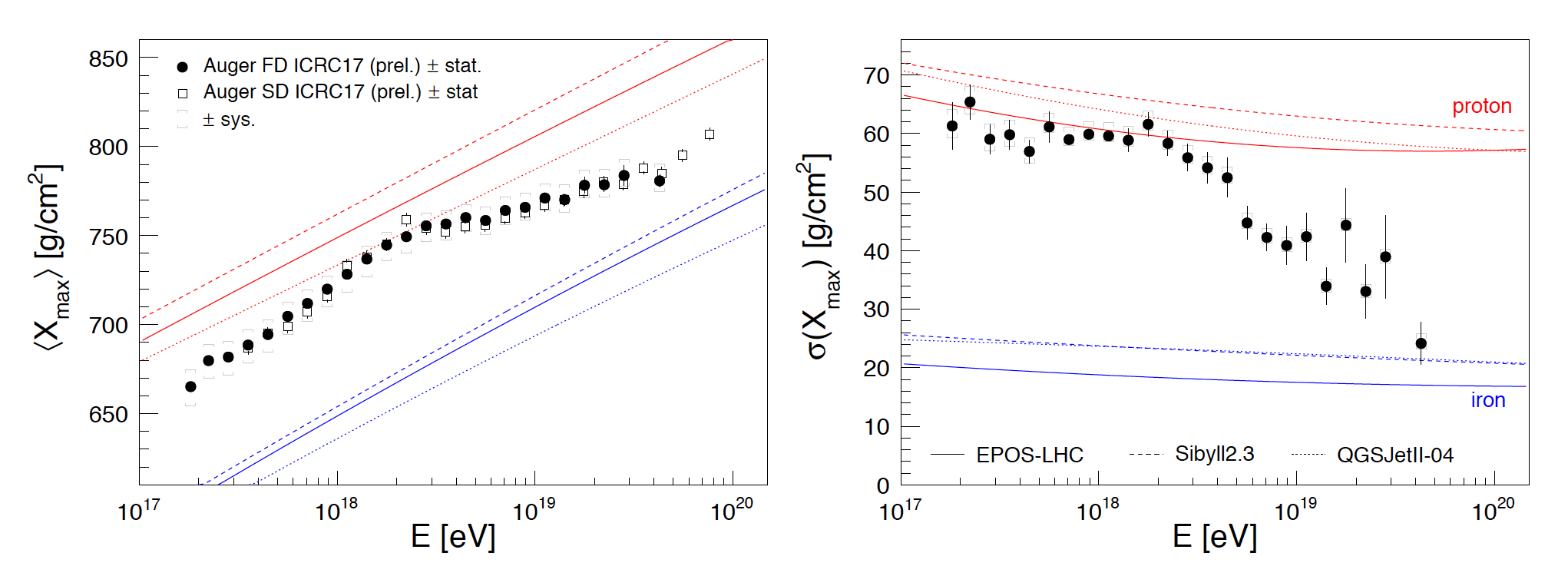}
\end{center}
\caption{The mean (left) and the standard deviation (right) of the $X_\mathrm{max}$
distributions measured by Auger, as a function of energy compared to air-shower simulations for protons and iron primaries.}
\label{fig:AugerXmax}
\end{figure*}

\begin{figure*}[t]
\begin{center}
 \includegraphics[scale=0.35]{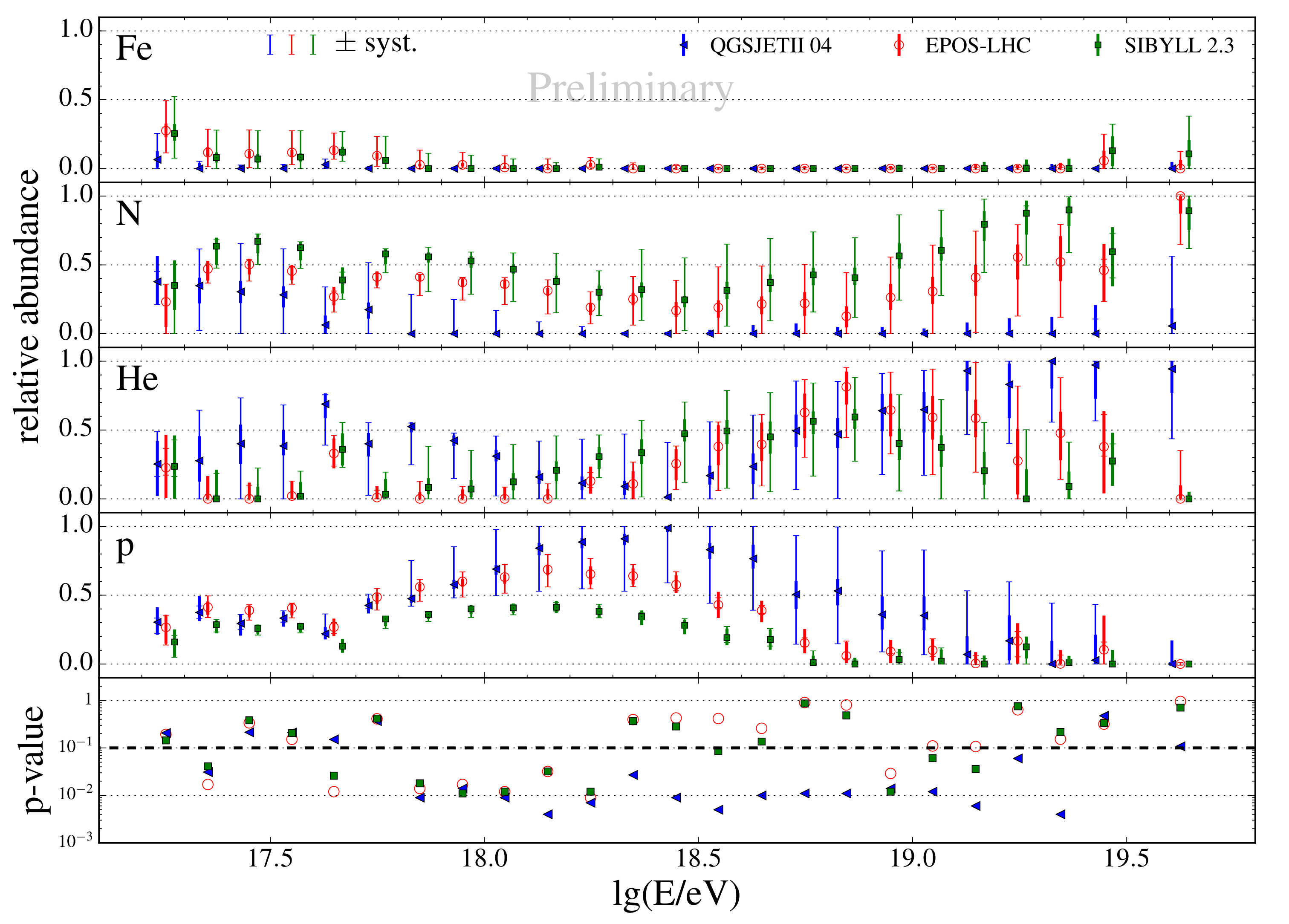}
\end{center}
\caption{Results from a fit of the $X_\mathrm{max}$ distributions with a superposition of H, He, N and Fe induced air
showers. The error bars indicate the statistics (smaller cap) and the systematic
uncertainties (larger cap). The bottom panel indicates the goodness of the fits ($p$-values).}
\label{fig:AugerFrac}
\end{figure*}
\section{Mass composition}
\label{sec:mass}
Composition is addressed using the depth of the position of the maximum in the energy deposit of shower particles, $X_\mathrm{max}$, which is measured by the FD. In a simplistic picture, the sensitivity of $X_\mathrm{max}$ to mass composition relies on the fact that showers from heavier (lighter) nuclei develop higher (deeper) in the atmosphere and their profiles fluctuate less (more). 

The measurements by Auger are robust for the accurate data selection and the statistical quality of the $X_\mathrm{max}$ distributions that are obtained. For the limited field of view of the telescopes, depending on the zenith angle and impact point of the shower, a fluorescence detector views a different range of $X_\mathrm{max}$. The Auger analysis adopts event selection and quality cuts that allow us to get rid of this bias and thus obtain unbiased $X_\mathrm{max}$ distributions. Correcting for detector resolution and acceptance, the first two moments of the distributions (mean and standard deviations) can be directly compared to air shower simulations. The Auger Collaboration has published $X_\mathrm{max}$ measurements for hybrid showers having energies above 10$^{17.8}$ eV~\cite{Aab:2014kda} and recently reported preliminary results extending these measurements down to 10$^{17.2}$ eV~\cite{Aab:2017njo,bellidoICRC17}.  Fig. \ref{fig:AugerXmax} shows the latest data. In terms of average mass cosmic rays evolve towards a lighter composition between 10$^{17.2}$ and 10$^{18.3}$ eV, qualitatively corresponding to a transition from a heavy Galactic composition to a light
extragalactic composition. At higher energies the trend is reversed and the average mass increases with energy.

The comparison of the $\langle X_\mathrm{max} \rangle$ energy dependence between
Auger and TA is not immediate because different approaches are used to measure this observable by each experiment. In a report by the joint Auger-TA Working Group \cite {Hanlon:2018dhz} 
methods to facilitate comparison of $\langle X_\mathrm{max} \rangle$ measurements are presented. Using these methods the Auger and TA composition results are shown to agree within the systematic uncertainties quoted by the two experiments. A paper by W.~Hanlon in these proceedings also addresses
this comparison study.

The statistics collected with the relatively low duty cycle of FD do not yet allow us to study the composition at energies where the flux suppression is observed. Using SD observables, it is however possible to cover also the highest
energy range. In ref. \cite{Aab:2017cgk} the time profiles of the signals
recorded with the SD stations are employed to derive risetime. This observable depends
on the distance of the shower maximum to ground and the relative amount of muons and electrons
detected. The risetime-related variable $\Delta_s$ correlates with
$X_\mathrm{max}$ and this allows to calibrate $\Delta_s$ to $\langle X_\mathrm{max} \rangle$. The energy evolution of $\langle X_\mathrm{max} \rangle$ from the SD is shown superimposed to the
one from the FD measurements in the left panel of Fig. \ref{fig:AugerXmax}. As can be seen, the two measurements are in good agreement, as is to be expected due to the cross-calibration. At ultrahigh energies the superior statistics of the SD give two more data points. More and other analysis methods exploiting other SD-based observables will in future complement this method and provide a better understanding of this energy region.

The Auger $X_\mathrm{max}$ data (moments and distributions) enable a step further in the interpretation of mass composition studying the evolution with energy of the first two moments of $\ln A$~\cite{Aab:2014kda,bellidoICRC17,Aab:2017njo} and of the fractions of four mass groups (H, He, N and Fe) from the fit of the $X_\mathrm{max}$ distributions~\cite{bellidoICRC17,Aab:2014aea,Aab:2017njo}. The latest results are shown in Fig. \ref{fig:AugerFrac}.
At the lowest energies, we find hints for a contribution (25$\div$38\% depending on  models) from iron primaries that disappears rapidly with increasing energy. At high energies
the composition is dominated by different elemental group, starting from protons below the ankle and going
through helium to nitrogen as the energy increases. This evolution occurs with limited mass mixing as
a consequence of the small values of the $X_\mathrm{max}$ dispersion. The interpretation of $X_\mathrm{max}$
data depends on the hadronic interaction models. In particular QGSJetII-04 appears to be less consistent with data as can be seen in the lower panel of Fig. \ref{fig:AugerFrac}, where the probability of the
fits is shown. 

\section{Astrophysical implications from spectrum and composition data}
\label{sec:impli}
The results shown in the previous section show our current knowledge of the mass composition of
UHECRs as they hit the Earth. In Sec. \ref{sec:spectrum} I pointed out that despite the accurate
measurement of the energy spectrum an interpretation in terms of
sources is ambiguous. Instead using both spectrum and composition one can remove this degeneracy and 
infer information about the source scenarios which are compatible with data.
Several investigations  have been done in recent years to interpret UHECR spectrum and composition \cite{Aloisio:2017ooo}. Most of these studies converge to scenarios with sources injecting hard spectra with low
rigidity cutoff and mixed composition, even though simplifying assumptions are used as uniform
source distributions and 1D cosmic ray propagation. These studies mainly address the description
of the data above the ankle because lower energy data need additional hypotheses as new source populations  or interactions of ejected cosmic rays in the radiation field surrounding the source \cite{Unger:2015laa,Globus:2015xga}. All these results are amply model dependent \cite{Batista:2015mea}: besides the
hadronic interaction models which describe the shower development in the atmosphere, the other model
uncertainties come from the extra-galactic background light (EBL) radiation which cosmic rays cross in
their propagation and the cross sections of photo-disintegration of nuclei interacting with background photons. These uncertainties are sizeable and mainly due to the lack of data \cite{Boncioli:2016lkt}.

\begin{figure}[!t]
\centering
\includegraphics*[scale=0.33]{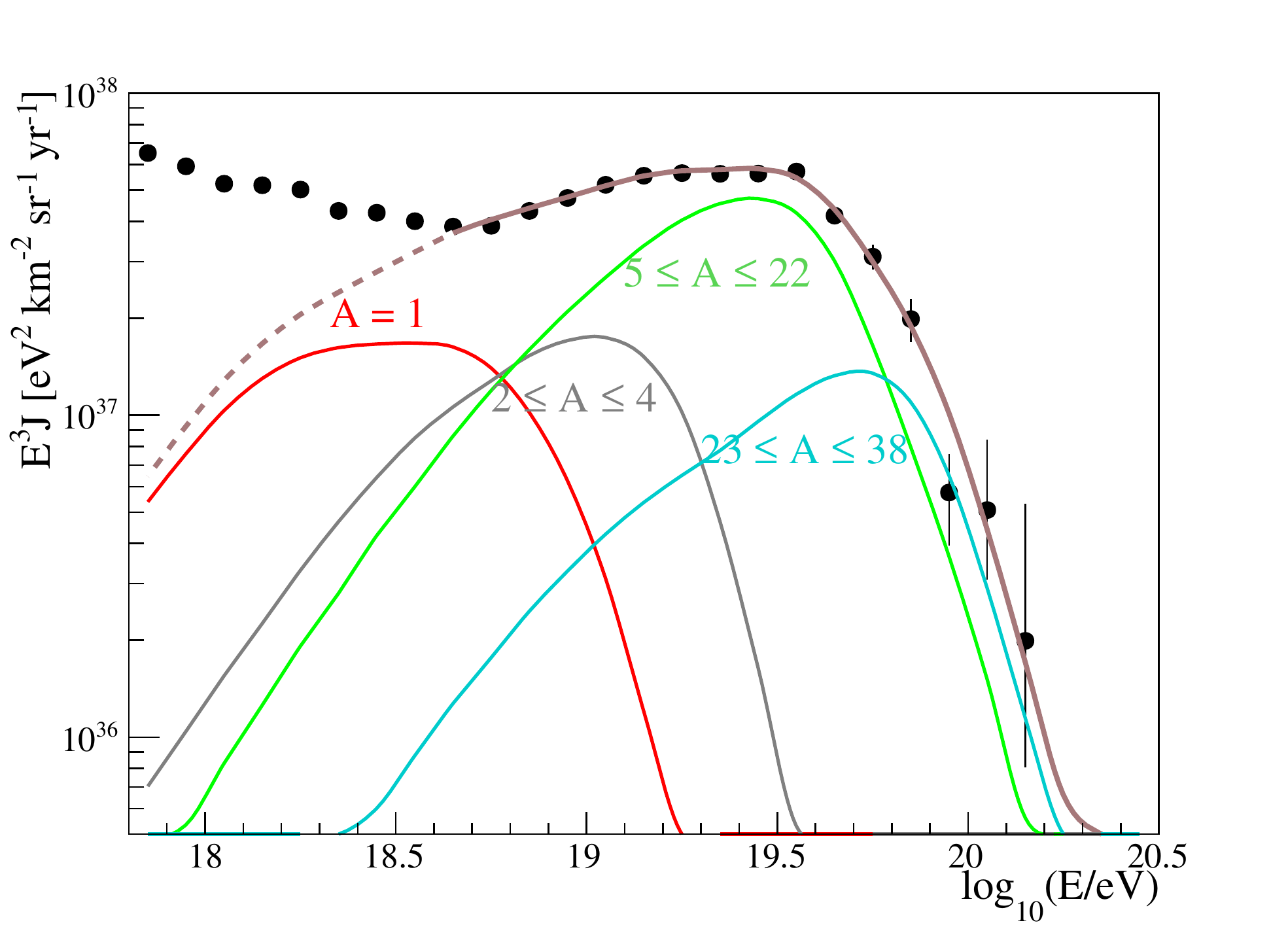}\\
    \includegraphics[scale=0.42]{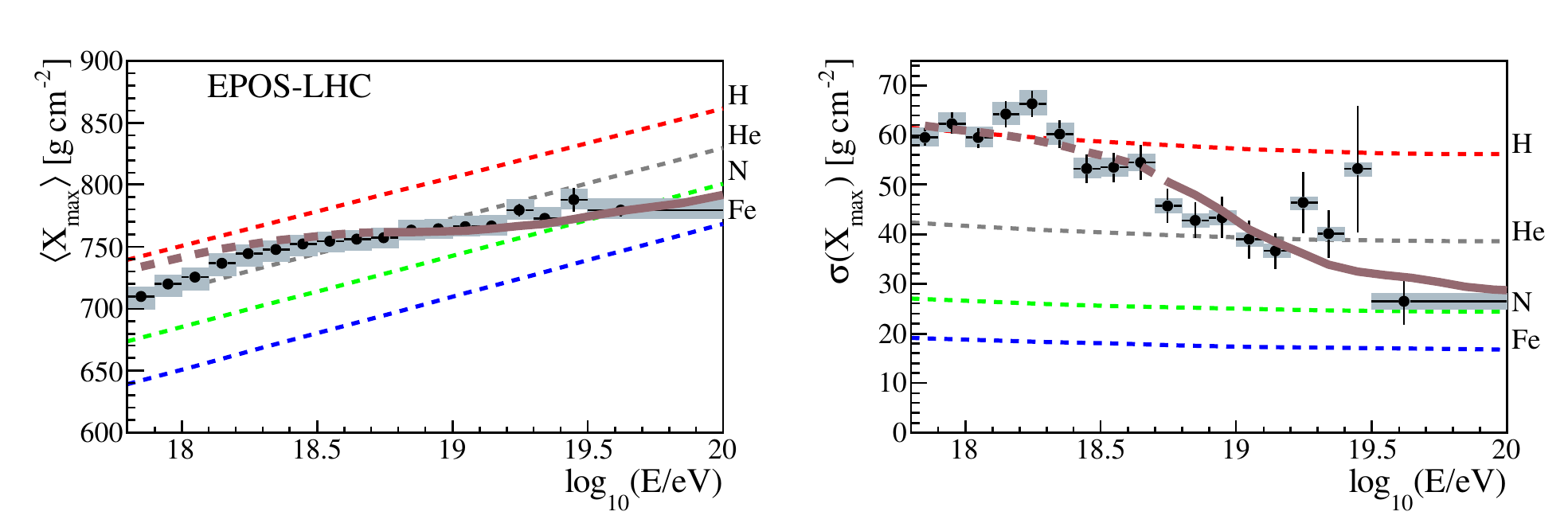}
    \caption{{\it Top:} energy spectrum  obtained with the best-fit parameters
      for the SPG model (SimProp code, PBS photo-production cross-sections,  Gilmore '12 EBL) \cite{refModels}. Partial spectra are shown with different colors, total spectrum with brown. 
 Full circles show the ICRC 2015 Auger spectrum.
 {\it Bottom:} average and standard deviation  of the $X_\mathrm{max}$ for data \cite{Aab:2014kda}
 (full circles) and best-fit prediction using EPOS-LHC (brown). Pure H (red), He (grey), N (green) and Fe (blue) are shown as dashed lines. Only the energy range where the brown lines are solid is included in the fit.}
\label{fig:SPGfit}
\end{figure}
The Auger Collaboration has published a comprehensive study about the astrophysical implications from the combined fit of 
spectrum and composition data \cite{Aab:2016zth},  discussing in detail the effects of theoretical
uncertainties on propagation and interactions in the atmosphere of UHECRs as well as the dependence of the 
fit parameters on the experimental systematic uncertainties. In this study we used a scenario
in which the sources of UHECRs are of extragalactic origin
and accelerate nuclei in electromagnetic processes with a rigidity-dependent maximum energy,
$E_\mathrm{max}(Z)=E_\mathrm{max}(p) \times Z$, where $Z$ denotes the charge and $E_\mathrm{max}(p)$ is the maximum energy for protons. Within this scenario a good description of the shape of
the measured energy spectrum as well as the energy evolution of the $X_\mathrm{max}$ distributions can be
achieved if the sources accelerate a primary nuclear mix consisting of H, He, N and Si, if the
primary spectrum follows a power law $\propto E^{-\gamma}$ with a spectral index $\gamma \approx 1$ and if the maximum energy of protons is about 10$^{18.7}$ eV, as shown in Fig. \ref{fig:SPGfit}. 
The mass composition at the sources is dominated by intermediate mass nuclei (N, Si). Using different hypotheses (i.e. hadronic interaction and
EBL models, photo-disintegration cross sections) \cite{refModels} source
parameters can change considerably and for some of them well outside the
statistical uncertainties of the fit with other assumptions. However the low spectral
index and low rigidity solution is generally preferred.

\begin{figure}[t]
\begin{center}
 \includegraphics[scale=0.32]{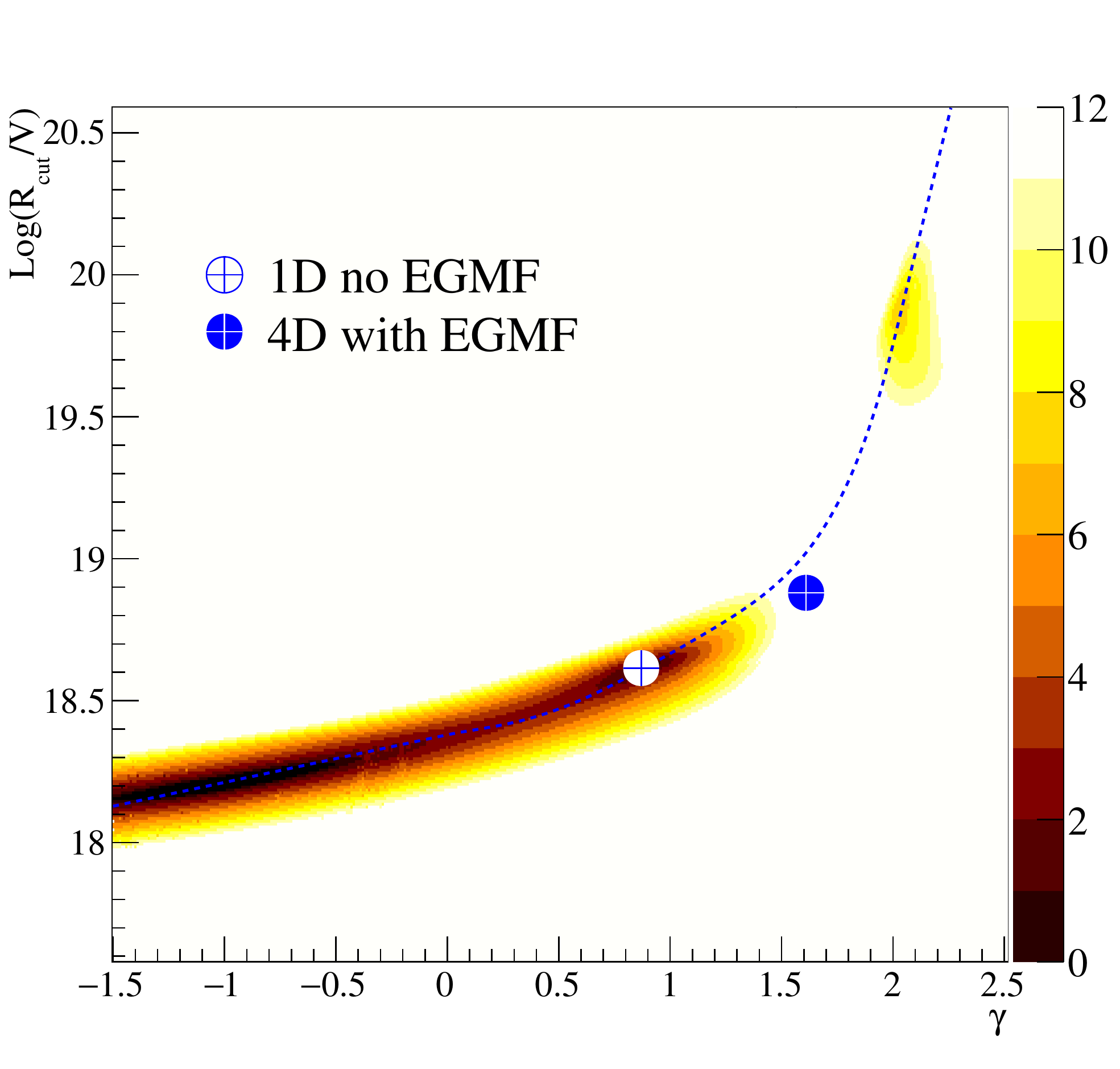}
\end{center}
\caption{Pseudo standard deviation $\sqrt{D - D_\mathrm{min}}$ (where $D$ is the deviance) as function of $\gamma$ and
    $\log_{10}(R_\mathrm{cut}/\mathrm{V})$ for the 1D propagation fit \cite{Aab:2016zth}. The filled blue (white) marker shows the position of the minimum for 4D propagation with LSS sources and EGMF (1D
  propagation).}
\label{fig:BestFit-1D-4D}
\end{figure}

In a more recent work \cite{Wittkowski:2017okb,Aab:2017njo} the homogeneous distribution of sources assumed in our previous
calculations was replaced by discrete sources distributed according to the model of the local
large-scale structure from \cite{Dolag:2004kp}, with a source density of $10^{-4}$ Mpc$^{-3}$.
Moreover, we studied the effects of the extragalactic magnetic field (EGMF) using the EGMF
model proposed in \cite{Batista:2016yrx}, which describes a relatively strong magnetic field, together with reflective boundary conditions.
To pursue this program the 4D mode
of the Monte Carlo code CRPropa 3 \cite{Batista:2016yrx} was used. The combined fit of spectrum and
composition provides  a good overall
description of the data (similar to the one achieved with the simpler model). Comparing
the best-fit parameters of the extended model with our previous results, we found that the details
of the local large-scale structure are of minor importance for the derived parameters of the source
spectra. Yet by including the diffusion in the EGMF in the calculation we derive a spectral index
of $\gamma \sim 1.6$, i.e. significantly softer than the one obtained without magnetic fields.
Fig. \ref{fig:BestFit-1D-4D} shows the change of the spectral parameters from the 1D (uniform, no EGMF) to the 4D (LSS source distribution + EGMF). The presence of magnetic fields in the intergalactic space needs therefore be taken into account when interpreting cosmic ray data, especially if the field strength is relatively strong as assumed in this study.

\section{Anisotropy} 
\label{sec:anisotropy}
The Auger Collaboration has undertaken several anisotropy searches at different energy ranges and angular scales. These use several tools like harmonic analysis, auto-correlation, correlation with source catalogs, search for flux excesses  in the visible sky and correlations with other experiments.

\begin{figure}[t]
\begin{center}
\includegraphics[scale=0.22]{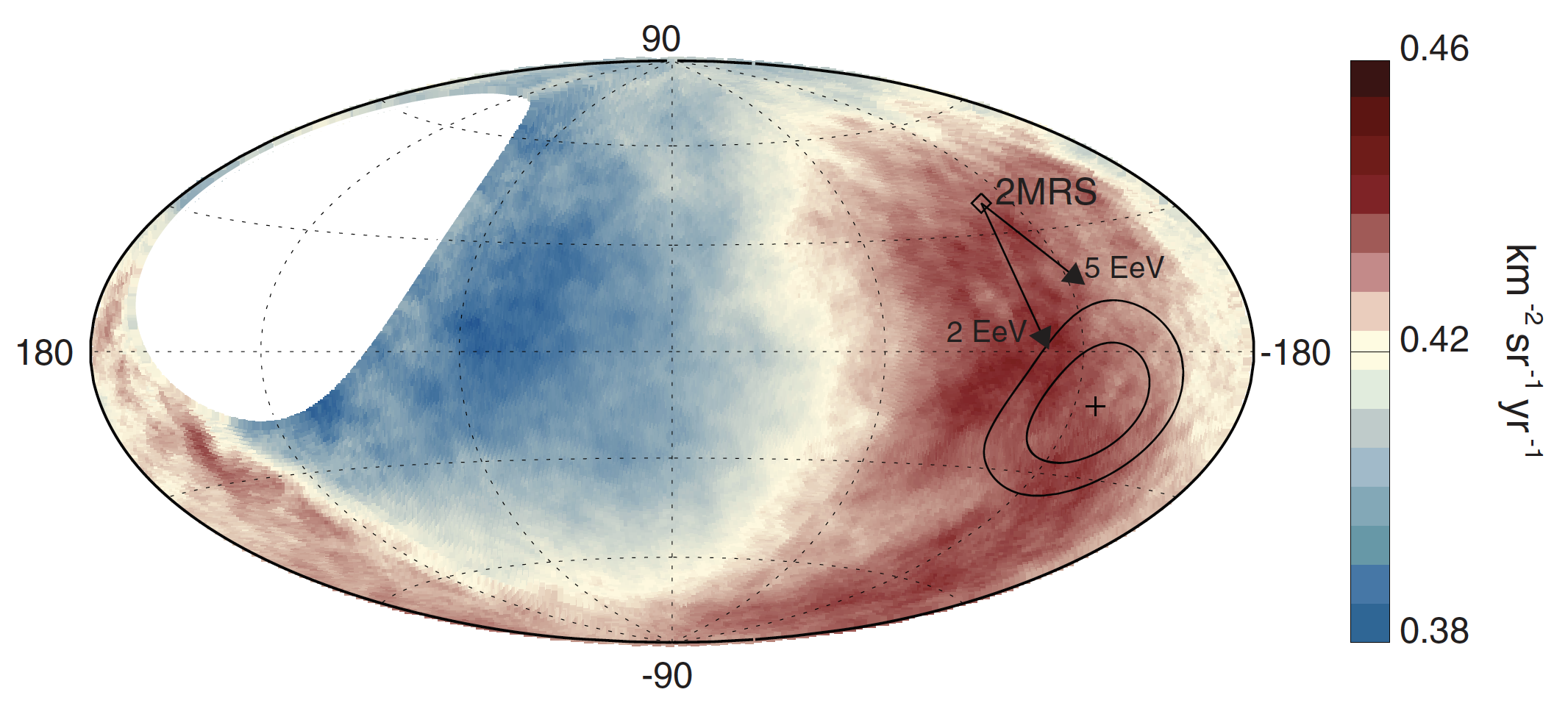}
\end{center}
\caption{Sky map in galactic
coordinates showing the cosmic-ray flux for $E \geq 8$ EeV smoothed with a 45$^\circ$ top-hat function. The
galactic center is at the origin. The cross indicates the measured dipole direction; the contours
denote the 68\% and 95\% confidence level regions. The dipole in the 2MRS galaxy distribution is
indicated. Arrows show the deflections expected for a particular model of the galactic magnetic
field \cite{Jansson:2012pc} on particles with $E/Z$ = 5 or 2 EeV.}
\label{fig:dipole8EeV}
\end{figure}

Among the most recent studies the  observation of a large-scale
anisotropy in the arrival directions of cosmic rays above $8 \times 10^{18}$ eV \cite{Aab:2017tyv} is indeed the most exciting. Two energy bins, 4 EeV $< E <$ 8 EeV and $E \geq 8$ EeV, were analysed
since the start of data taking (total exposure of 76,800 km$^2$ sr yr) with the Observatory  measuring the amplitude of the first harmonic
in right ascension. The right ascension anisotropy found in the two energy bins has amplitude  $0.5^{+0.6}_{-0.2}\%$ and
  $4.7^{+0.8}_{-0.7}\%$, respectively. The events in the lower energy bin
follow an arrival distribution consistent with isotropy, but in the higher energy bin a significant
anisotropy was found, with a $p$-value of $2.6 \times 10^{-8}$ under the isotropic null hypothesis.
 The three-dimensional dipole, obtained combining the
first-harmonic analysis in right ascension with a similar one in the azimuthal angle, has a direction in
galactic coordinates
$(l, b) = (233^\circ, -13^\circ)$ about 125$^\circ$ away from the Galactic Centre hence indicating an extragalactic origin for these UHECR particles. The dipole anisotropy is detected at  more than a 5.2$\sigma$ level of significance with an amplitude of $6.5^{+1.3}_{-0.9}\%$. A skymap of the intensity of cosmic rays arriving above 8 EeV is shown in Fig. \ref{fig:dipole8EeV}.

 Some large scale anisotropy is expected because of the relative motion of cosmic rays with respect to the rest frame of background radiation~\cite{Kachelriess:2006aq} but the amplitude is expected to be below percent level, well below what has been observed. Other
 studies have predicted larger anisotropies originating from an inhomogeneous distribution of sources, or
 that they arise from a dominant source. To illustrate this effect we considered the distribution of nearby galaxies,
 as mapped by the 2MASS redshift survey (2MRS) \cite{2MRSCatalog}, which exhibits a dipolar structure. 
 Fig.  \ref{fig:dipole8EeV} shows the direction toward the 2MRS dipole  and the expected deflections
 caused by the galactic magnetic field. It is worth noting that the agreement between the directions of the
 UHECR and 2MRS dipoles
 is improved by adopting assumptions consistent with the observed charge composition and the deflections in the Galactic
 magnetic field.

 An extension of the analysis of the large-scale anisotropy has been recently submitted \cite{Aab:2018mmi}. Here both the dipolar and quadrupolar components are studied in the two energy ranges and further 
 the bin above 8 EeV is analysed by splitting it into three so as to explore how the amplitude and phase of the dipole changes with energy. The quadrupolar component is found to be not statistically significant. Instead we find that the amplitude of the dipole increases with energy above 4 EeV, as expected by predictions from models \cite{Harari:2015hba}.

The search for flux excesses does not 
show statistically significant evidence of anisotropy \cite{PierreAuger:2014yba,Giaccari:2017cvc}. Yet remarkable flux excesses is observed at intermediate scales. The strongest departure from isotropy (post-trial significance $\sim 3.1 \sigma$) is obtained for cosmic rays with $E > 58$ EeV around the direction of Cen A (15$^\circ$ radius). It is remarkable  to recall that for cosmic ray events with energy $E > 57$ EeV, Telescope Array have observed in the northern sky a cluster of events (hotspot) \cite{Abbasi:2014lda}, centered at $\alpha_\mathrm{d} = 146.7^\circ$, $\delta_\mathrm{d} = 43.2^\circ$, of about 20$^\circ$ radius and with a calculated probability of appearing by chance in an isotropic cosmic-ray sky of 3.7$\times 10^{-4}~(3.4 \sigma)$. It will be interesting to follow the evolution of these excesses with future data from both experiments.

\begin{figure}[t]
\begin{center}
\includegraphics[scale=0.4]{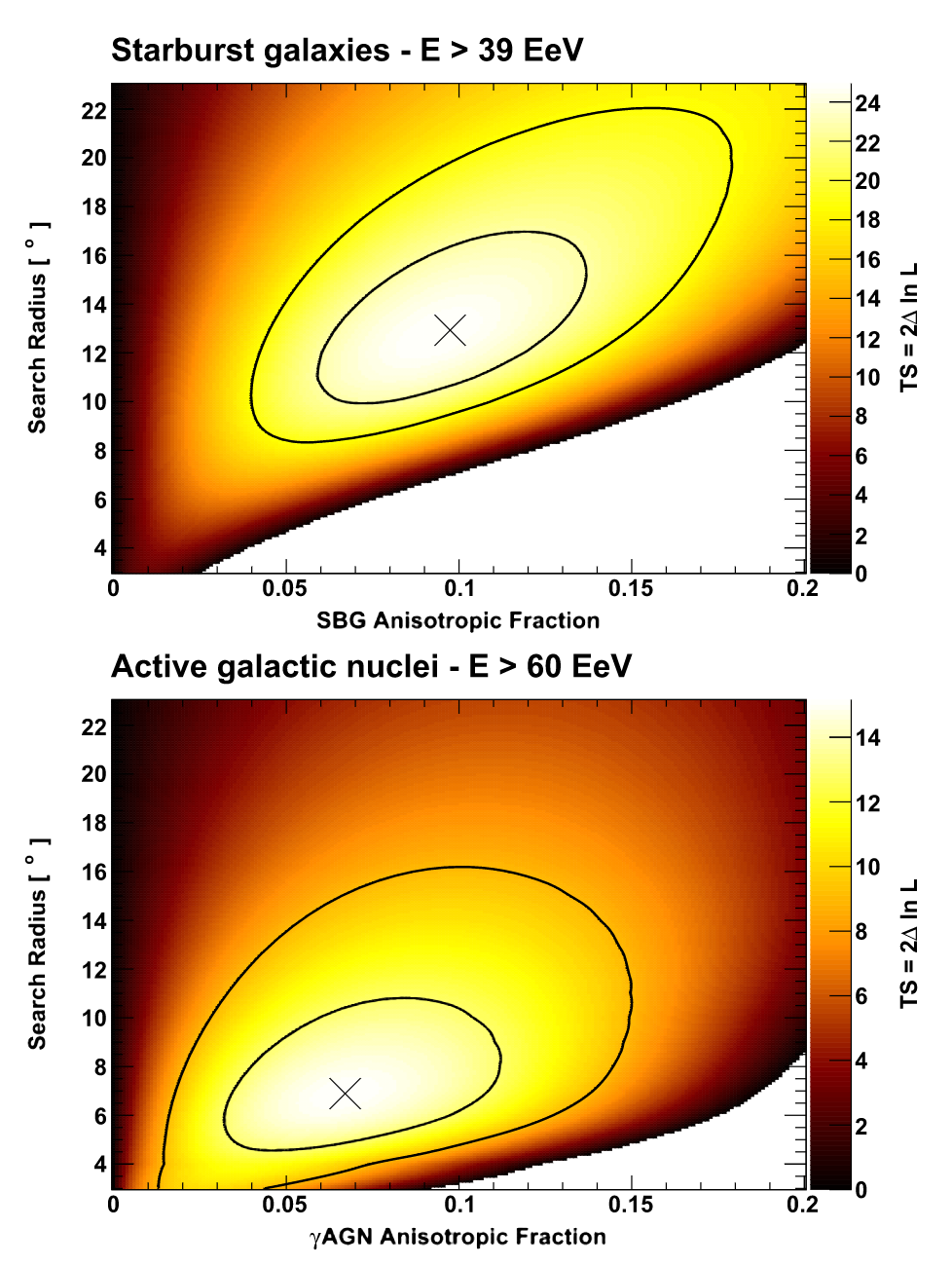}
\end{center}
\caption{TS profile above 39 EeV (top) and 60 EeV (bottom) over the fit parameters for SBG and $\gamma$AGN models . The lines indicate the 1$\sigma - 2\sigma$
regions.}
\label{fig:Best}
\end{figure}

A different type of analysis \cite{Aab:2018chp} has been also performed, based on the assumption that the UHECR flux is proportional to non-thermal electromagnetic flux. This analysis then takes into account the different flux of the single candidate sources. Two different candidate sources were taken into account: AGNs and starburst (starforming) galaxies (SBG). Active galaxies were extracted from the Fermi-LAT 2FHL Catalog \cite{Ackermann:2015uya}, selecting only radio-loud AGNs within a 250 Mpc radius. A list of 17 bright nearby candidates was obtained this way, and their integral $\gamma$-ray fluxes between 50 GeV and 2 TeV were used as a proxy for the UHECR flux. For SBGs \cite{Gao:2003qp}, since only a few of them were observed directly
in the $\gamma$-ray band, continuum emission at 1.4 GHz was used
as a proxy for the UHECR flux so that the 23 brightest nearby objects with a radio flux larger than 0.3 Jy were selected.
\begin{figure}[t]
\begin{center}
\includegraphics[scale=0.28]{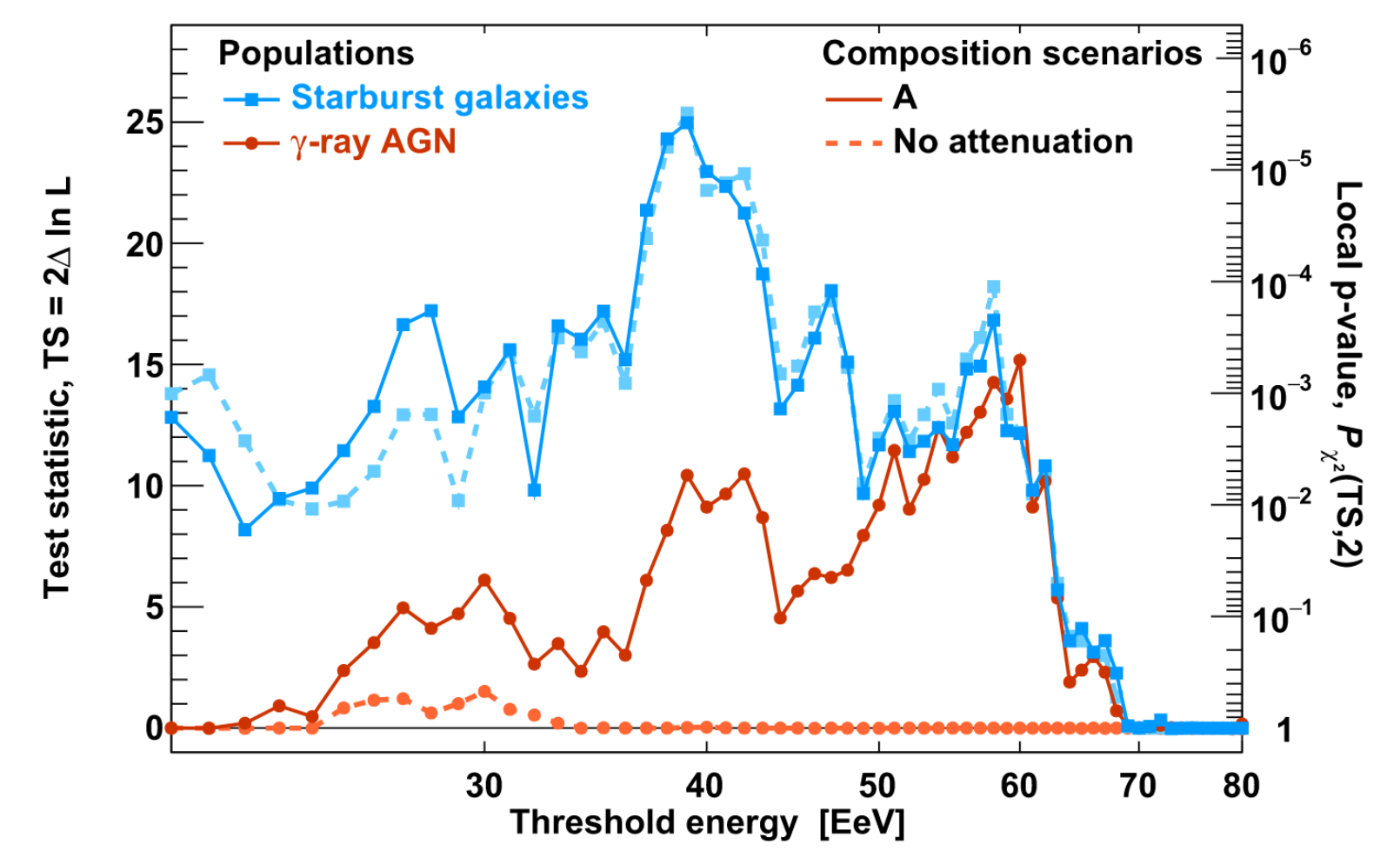}
\end{center}
\caption{Test statistic (TS) scan over the threshold energy for SBGs and $\gamma$-AGNs.
  The continuous lines indicate the values of the test
statistics obtained accounting for attenuation of the intensity due to energy losses while the dotted lines refer to the values without any attenuation.}
\label{fig:TS}
\end{figure}
The attenuation
of the intensity due to energy losses in the propagation to Earth is taken into account. The  free parameters of this study are the threshold energy, the smearing angle and the fraction of anisotropic cosmic rays originating from the tested intensity model. The evolution of the test statistic (the likelihood ratio between model and isotropy) as a function of threshold energy
is shown in Fig. \ref{fig:TS}.
The TS is then maximized
as a function of two free parameters in different energy threshold from 20 up to 80 EeV. In Figure \ref{fig:Best}, the TS as a function of the two free parameters is shown for the energy threshold where the
maximum is found: 60 EeV for the AGNs and 39 EeV for the starburst galaxies. The best-fit
parameters are found to be a smearing angle of 13$^\circ$ and an anisotropic fraction of 10\% for the
starburst-galaxies with a TS of 24.9, corresponding to a significance of $\sim 4 \sigma$, and 7$^\circ$ and 7\% for
the $\gamma$-ray AGNs with a TS of 15.2 corresponding to a significance of $\sim 2.7 \sigma$. It is remarkable
that while the significance for the AGNs is close to the one obtained in a previous analysis \cite{PierreAuger:2014yba}, a much higher significance is obtained with the newly tested starburst hypotesis.

\section{Future prospects: AugerPrime}
UHECR data provide several interesting outcomes on several aspects of cosmic ray physics as the ones presented in this work, but also others including photon and neutrino limits, multi-messengers studies and hadronic interactions at energies higher than LHC,  not discussed here. Despite the progresses in our understanding it appears still difficult to build a consistent picture of the origin of UHECRs. To make further progress in this direction more accurate and extended information on the nature of the primaries is required: mass composition is currently unavailable above 40 EeV due to the intrinsic duty cycle of the FD and the scarce accuracy of the composition sensitive methods based on the surface array data. 

The  AugerPrime  upgrade~\cite{Aab:2016vlz}  of  the  Pierre Auger Observatory  has  been  specifically  designed  to  improve  mass composition in the whole energy range.  Along  the  line of a hybrid design, each SD will be equipped  with  a  top  scintillator  layer. Shower  particles  will  be  sampled  by  two detectors  (scintillators and water-Cherenkov  stations) having different responses to the muonic and electromagnetic components. The combined measurement
allows to disentangle these two components and to provide an estimate of both, mass and energy
of the shower, on an event-by-event basis.
The detectors will be read out by new electronics with
a faster and more accurate sampling of the signal. An extra small photomultiplier installed in each WCD will extend the current dynamic range to more than 32 times the largest signals currently measured. The upgraded array will provide data with no duty cycle limitation and then the access to the highest energies will be made possible.
This setup is complemented by an underground muon detector AMIGA \cite{PierreAugur:2016fvp}, in the current infill array. Finally we plan to
increase the current FD duty cycle by $\sim 50\%$ extending the operational mode to periods with a higher night sky
background.

\begin{figure}[t]
\begin{center}
\includegraphics[scale=0.23]{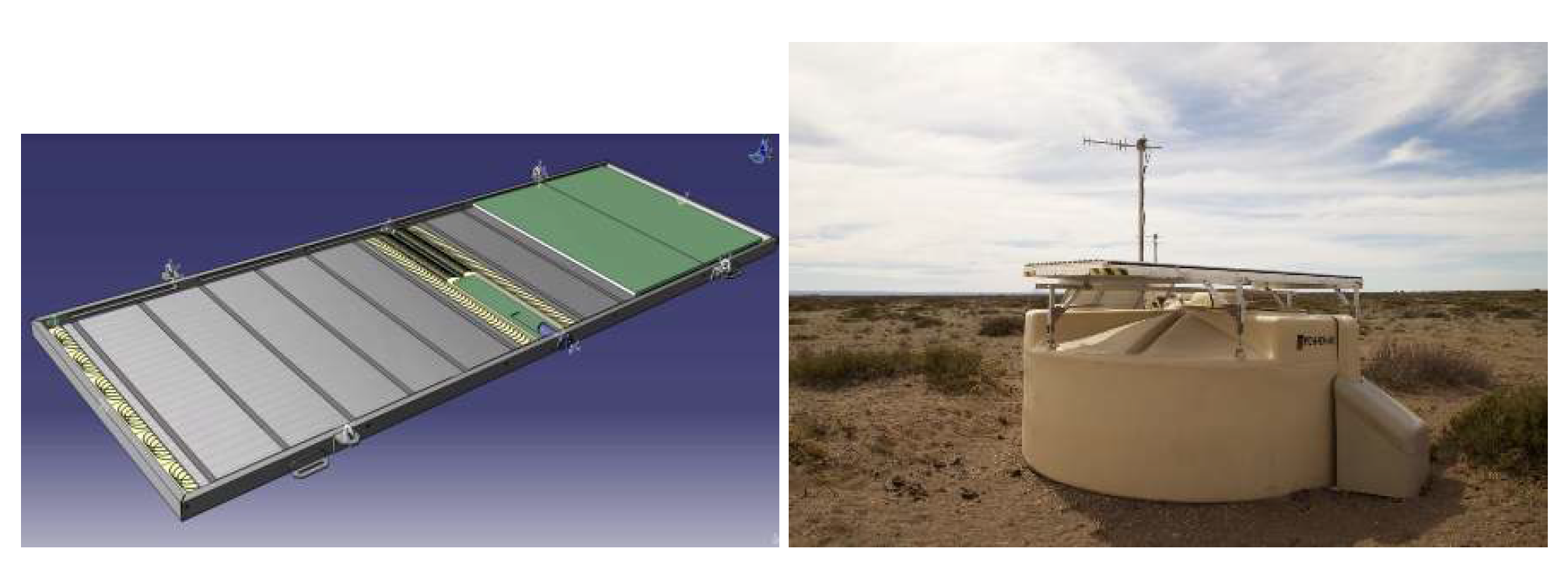}
\end{center}
\caption{{\it Left:} the layout of the Surface Scintillator Detector (SSD). {\it Right:} One station of the AugerPrime
Engineering Array.}
\label{fig:APdet}
\end{figure}

\begin{figure}[t]
\begin{center}
\includegraphics[scale=0.24]{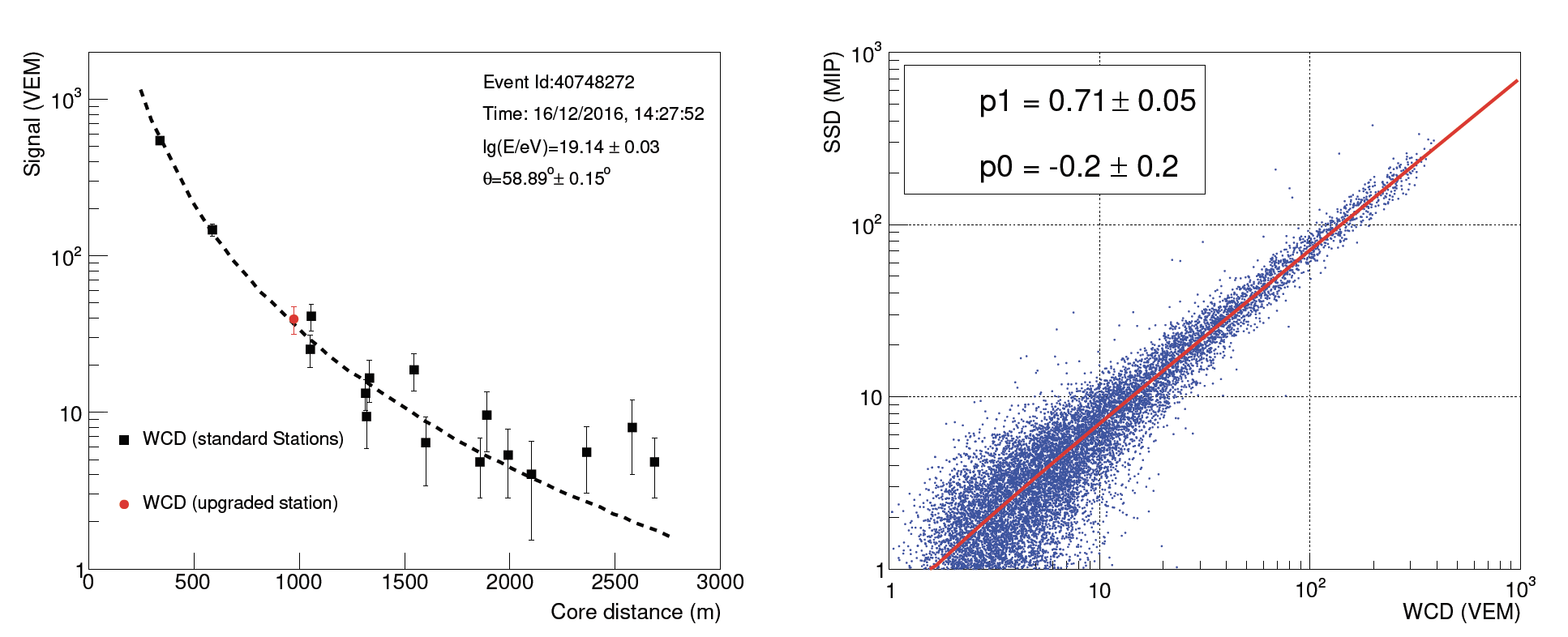}
\end{center}
\caption{{\it Left:} one event reconstructed with the regular 1500 m array in
  close proximity to the EA. The reconstructed signals in the EA are compared
  with the LDF of the event. {\it Right:} correlation of the signals of
the SSD and the WCD. Both signals have been calibrated.}
\label{fig:APdata}
\end{figure}
The AugerPrime Engineering Array (EA) \cite{Martello:2017pch,Aab:2017njo}
consisting of 12 AugerPrime detector stations is in operation since 2016.
The layout of the Surface Scintillator Detector
(SSD) stations is shown in Fig. \ref{fig:APdet} together with the picture
of one upgraded station of the EA.
With this setup we have verified the basic functionality of the detector design, the linearity of the scintillator signal, the calibration procedures and operational stability. The upgraded stations produce signals
that are in good agreement with expectations as shown in Fig. \ref{fig:APdata} for one event reconstructed
from EA data.

The construction of AugerPrime is expected to be finished by 2019 and it will take data until 2025.

\begin{acknowledgement}
\section*{Acknowledgements}
The successful installation, commissioning, and operation of the Pierre Auger Observatory would not have been possible without the strong commitment and effort from the technical and administrative staff in Malarg\"ue, and the financial support from a number of funding agencies
in the participating countries, listed at
{\href{https://www.auger.org/index.php/about-us/funding-agencies}{https://www.auger.org/index.php/about-us/funding-agencies}}.
\end{acknowledgement}

\end{document}